\documentclass[11pt]{article}
\usepackage{hyperref}
\pdfoutput=1

\begin{document}
\title{Acoustophoretic Waltz: \\a Contactless Exothermal Reaction}
\author{Daniele Foresti and Dimos Poulikakos\\
\\\vspace{6pt} Department of Mechanical and Process Engineering,
\\ ETH Zurich, Zurich, 8092, Switzerland}
\maketitle

\begin{abstract}
The fluid dynamics video shows the acoustophoretic handling of a metal sodium chunks and a water droplets before, during and after mixing. The violent exothermal reaction between solid and liquid introduces an additional phase (hydrogen gas). We developed a unique concept for using ultrasound to stably levitated and move along a plane multiple objects in air, independently from their electromagnetic nature and aspect ratio. This contactless material handling can be extended to hazardous, chemical or radioactive samples.
\end{abstract}

\end{document}